\documentclass{PoS}

\title{Understanding the properties of $\Xi(1690)$ and $\Xi(2120)$}

\ShortTitle{Understanding the properties of $\Xi(1690)$ and $\Xi(2120)$}

\author{\speaker{K. P. Khemchandani}$~^a$, A.~Mart\'inez~Torres$^b$,  A.~Hosaka$^c$, H.~Nagahiro$^{c,d}$, F.~S.~Navarra$^b$, M.~ Nielsen$^b$
\\  \llap{$öa$}
        Departamento de Ci\^encias Exatas e da Terra, Universidade Federal de S\~ao Paulo, Campus Diadema, Rua Prof. Artur Riedel, 275, Jd. Eldorado, 09972-270, Diadema, SP, Brazil.\\
   \llap{$öb$}  Instituto de F\'isica, Universidade de S\~ao Paulo, C.P 66318, 05314-970 S\~ao Paulo, SP, Brazil.\\
 \llap{$öc$} Research Center for Nuclear Physics (RCNP), Mihogaoka 10-1, Ibaraki 567-0047, Japan.\\
 \llap{$öd$} Department of Physics, Nara Women's University,  Nara 630-8506, Japan.

        E-mail: \email{kanchan.khemchandani@unifesp.br}, \email{amartine@if.usp.br}, \email{hosaka@rcnp.osaka-u.ac.jp}, \email{nagahiro@rcnp.osaka-u.ac.jp}, \email{navarra@if.usp.br}, \email{mnielsen@if.usp.br}}

\abstract{In this manuscript we discuss the results of our recent study on $\Xi$ resonances.  We have studied meson-baryon interaction with strangeness $-2$ by solving scattering equations in a coupled channel approach and found that the dynamics gives rise to the development of two poles in the complex plane close to the real axis. The corresponding  narrow states have isospin half and spin half in one case and spin 3/2 in the other case. We find that the properties of the isospin and spin 1/2 state are strikingly similar to $\Xi(1690)$, while the state with spin 3/2 can be related to $\Xi(2120)$. We find that our results on $\Xi(1690)$ are in good agreement with the data available from the  BELLE and BABAR Collaborations. Our formalism consists of treating pseudoscalar-baryon and vector-baryon channels on an equal footing. Since the thresholds of such channels are distributed over a relatively larger energy range, we obtain all interaction kernels needed to solve the Bethe-Salpeter equation without resorting to any nonrelativistic approximation even though the amplitudes are calculated at low energies. }

\FullConference{XVII International Conference on Hadron Spectroscopy and Structure - Hadron2017\\
		25-29 September, 2017\\
		University of Salamanca, Salamanca, Spain}

\begin{document}

\section{Introduction}
The spectrum of multistrange baryons consists of  a few states whose properties like masses and widths, spin-parity are not well known \cite{pdg}, in several cases even the existence of the state itself is not clear. The reason behind such a deficit of information is the lack of data on multistrange baryons, which arises due to the poor statistics on their production cross sections obtained from collisions of nonstrange hadrons and due to the availability of few facilities of anti-kaon beams. However, the latest improvements in the analyzing techniques have encouraged extraction of information on $\Xi$ resonances from rare decay processes like  $\Lambda_c \to K^+ \bar K^0 \Lambda^0$ and $\Lambda_c \to K^+ K^- \Sigma^+$ \cite{belle, babar}. Besides, new experiments on productions of  $\Xi$'s are expected to be  scheduled at facilities like, JPARC  \cite{jparc}, J-lab \cite{clasx}, $\overline{\rm P}$ANDA \cite{panda}. 

In addition to the motives stated above, we find it encouraging to study strangeness $-2$ resonances since we can implement and test our  formalism on meson-baryon interactions. As we will discuss in more detail in the next section, our formalism consists of obtaining the lowest order amplitudes by relying on the hidden-local symmetry for the vector meson-baryon interactions. It was shown in our previous work \cite{vbvb} that the vector meson-baryon interaction Lagrangian leads to a contact interaction, apart from Yukawa-type vertices which can be used to calculate $t$-, $s$- and $u$-channel interactions. In Ref.~\cite{vbvb}, it was found that all these interactions give comparable contributions and, thus, all must be considered. We have also included the transition amplitudes between pseudoscalar meson- and vector meson-baryon consistently with the hidden-local symmetry (for more details, see Ref.~\cite{pbvb}). The pseudoscalar-baryon interaction amplitudes are obtained from the lowest order chiral Lagrangian \cite{ecker,pich,osetramos}. Our formalism has been used to study systems with other quantum numbers \cite{hyperons,more}. 

After giving some more details on the formalism, we will summarize the most important features of  the results found in Ref.~\cite{cascades}. As we will show, we find that the coupled channel meson-baryon interaction generates two resonances; both with isospin 1/2, one with spin 1/2 and other with spin 3/2. We find no isospin 3/2 state which would be exotic pentaquark-like states. 

\section{Formalism and Results}
The formalism consists of obtaining the vector meson-baryon interaction kernels from 
the Lagrangian \cite{vbvb}
\begin{eqnarray} \nonumber
&&\mathcal{L}_{\textrm VB}= -g \Biggl\{ \langle \bar{B} \gamma_\mu \left[ V_8^\mu, B \right] \rangle + \langle \bar{B} \gamma_\mu B \rangle  \langle  V_8^\mu \rangle  
\Biggr. + \frac{1}{4 M} \left( F \langle \bar{B} \sigma_{\mu\nu} \left[ V_8^{\mu\nu}, B \right] \rangle  + D \langle \bar{B} \sigma_{\mu\nu} \left\{ V_8^{\mu\nu}, B \right\} \rangle\right)\\
&& +  \Biggl.  \langle \bar{B} \gamma_\mu B \rangle  \langle  V_0^\mu \rangle  
+ \frac{ C_0}{4 M}  \langle \bar{B} \sigma_{\mu\nu}  V_0^{\mu\nu} B  \rangle  \Biggr\},  \label{vbb}
\end{eqnarray}
where  $\langle ... \rangle$ refers to an $SU(3)$ trace, the subscript $8$ ($0$) on the meson fields denotes the octet (singlet) part of their wave function (relevant in case of $\omega$ and $\phi$, for which we assume an ideal mixing). $V^{\mu\nu}$ represents the  tensor field of the vector mesons,
\begin{equation}
V^{\mu\nu} = \partial^{\mu} V^\nu - \partial^{\nu} V^\mu + ig \left[V^\mu, V^\nu \right], \label{tensor}
\end{equation}
and $V^\mu$ and $B$ denote the SU(3) matrices for the (physical) vector mesons and octet baryons. The values of the constants in Eq.~(\ref{vbb}) are $D$ = 2.4, $F$ = 0.82, and $C_0 = 3F - D$, which are chosen to reproduce the known anomalous magnetic couplings. As can be seen from Lagrangian of  Eq.~(\ref{vbb}), both vector-baryon-baryon vertices as well as a contact term can be obtained, the latter coming from the commutator part of the tensor field (Eq.~(\ref{tensor})). The vector-baryon-baryon vertices can be used to deduce $s$- and $u$- channel diagrams, 
\begin{figure}[ht!]
\begin{center}	
\includegraphics[width= 0.68\textwidth]{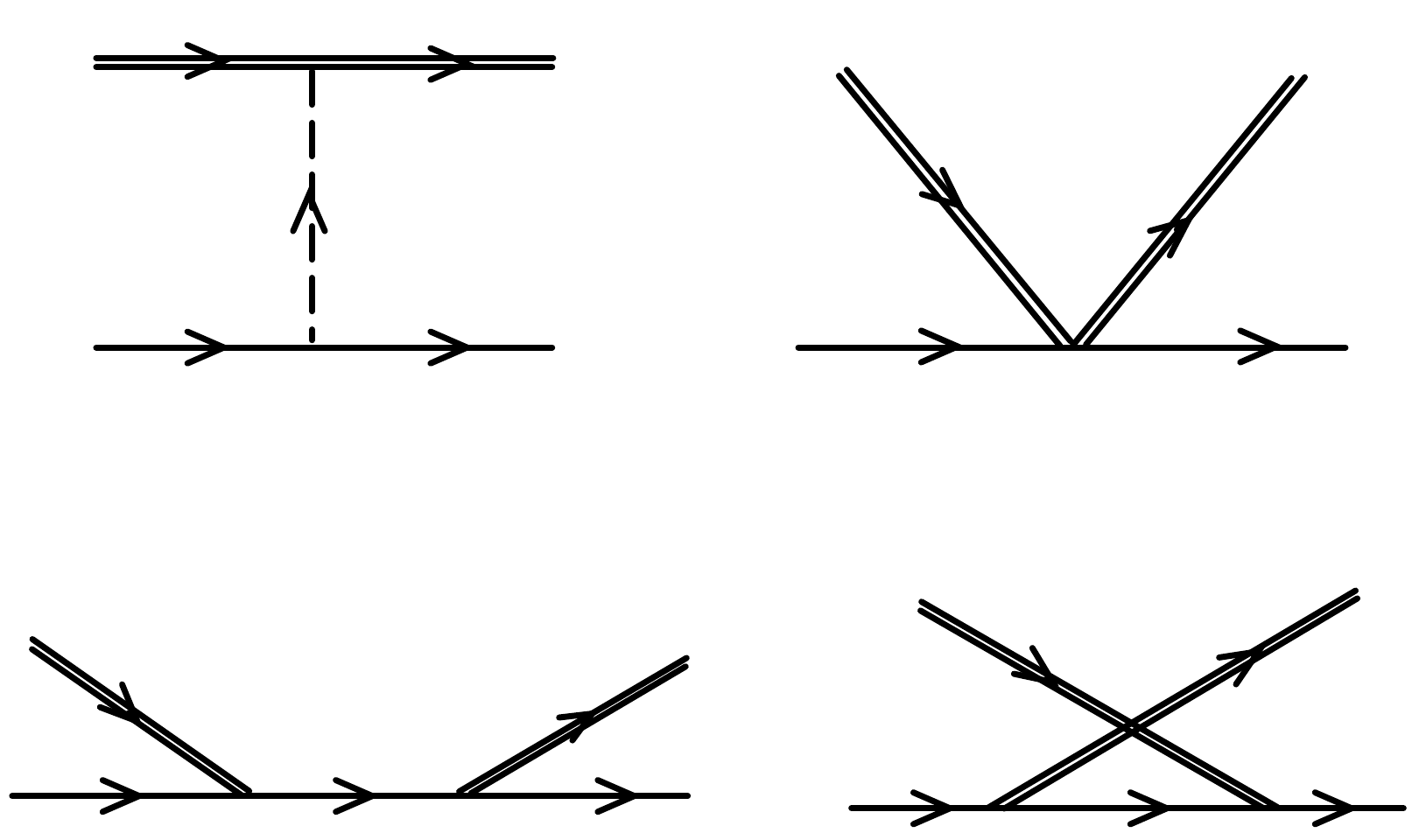}
\caption{Diagrams contributing to vector meson-baryon interactions. The double line in the figure represent vector mesons and single lines represent baryons. }\label{fig_vbinteractions}
\end{center}
\end{figure}
and $t$-channel diagrams can be calculated using the three-vector Lagrangian;
\begin{equation}
\mathcal{L}_{3V} \in - \frac{1}{2} \langle V^{\mu\nu} V_{\mu\nu} \rangle.
\end{equation}
The resulting relativistic amplitudes for the different coupled channels ($\pi \Xi$,  $\eta \Xi$, $\bar K \Sigma$, $\bar K \Lambda$, $\rho \Xi$,  $\omega \Xi$, $\phi \Xi$, $\bar K^* \Sigma$ and $\bar K^* \Lambda$) are given in Ref.~\cite{cascades}. The $t$-channel amplitudes obtained in our work, in the limit of nonrelativistic approximation, are in agreement with those in Ref.~\cite{ramosvb,Gamermann}. The sum of all the four diagrams shown in Fig.~\ref{fig_vbinteractions} are used as vector meson-baryon interaction kernel in the calculation of the Bethe-Salpeter equation.

For the pseudoscalar meson-baryon amplitudes, we rely on the lowest order chiral Lagrangian~\cite{ecker,pich,osetramos}
\begin{eqnarray}
\mathcal{L}_{PB} &=& \langle \bar B i \gamma^\mu \partial_\mu B  + \bar B i \gamma^\mu[ \Gamma_\mu, B] \rangle - M_{B} \langle \bar B B \rangle  
+  \frac{1}{2} D^\prime \langle \bar B \gamma^\mu \gamma_5 \{ u_\mu, B \} \rangle + \frac{1}{2} F^\prime \langle \bar B \gamma^\mu \gamma_5 [ u_\mu, B ] \rangle,~~\label{LPB}
\end{eqnarray}
where  
\begin{eqnarray}\nonumber
\Gamma_\mu &=& \frac{1}{2} \left( u^\dagger \partial_\mu u + u \partial_\mu u^\dagger  \right), \,u_\mu = i u^\dagger \partial_\mu U u^\dagger ,\,\\
U&=&u^2 = {\textrm exp} \left(i \frac{P}{f_P}\right),\label{gammau}
\end{eqnarray}
In the equations above, $f_P$ is the pseudoscalar decay constant,  $P$ represents the SU(3) matrix for the pseudoscalar mesons and  $F^\prime = 0.46$, $D^\prime = 0.8$ reproduce the axial coupling constant of the nucleon:  $F^\prime + D^\prime \simeq  g_A = 1.26$.

The  transition between pseudoscalar meson-baryon and vector meson-baryon channels is obtained by using the Lagrangian \cite{pbvb}
\begin{eqnarray} \label{pbvbeq}
\mathcal{L}_{\rm PBVB} &=& \frac{-i g_{KR}}{2 f_\pi} \left( F^\prime \langle \bar{B} \gamma_\mu \gamma_5 \left[ \left[ P, V^\mu \right], B \right] \rangle \right.
\left. D^\prime \langle \bar{B} \gamma_\mu \gamma_5 \left\{ \left[ P, V^\mu \right], B \right\}  \rangle \right).
\end{eqnarray}
 The amplitudes for different channels are given in detail in Ref.~\cite{cascades}.

\begin{figure}[ht!]
\begin{center}	
\includegraphics[width= 0.88\textwidth]{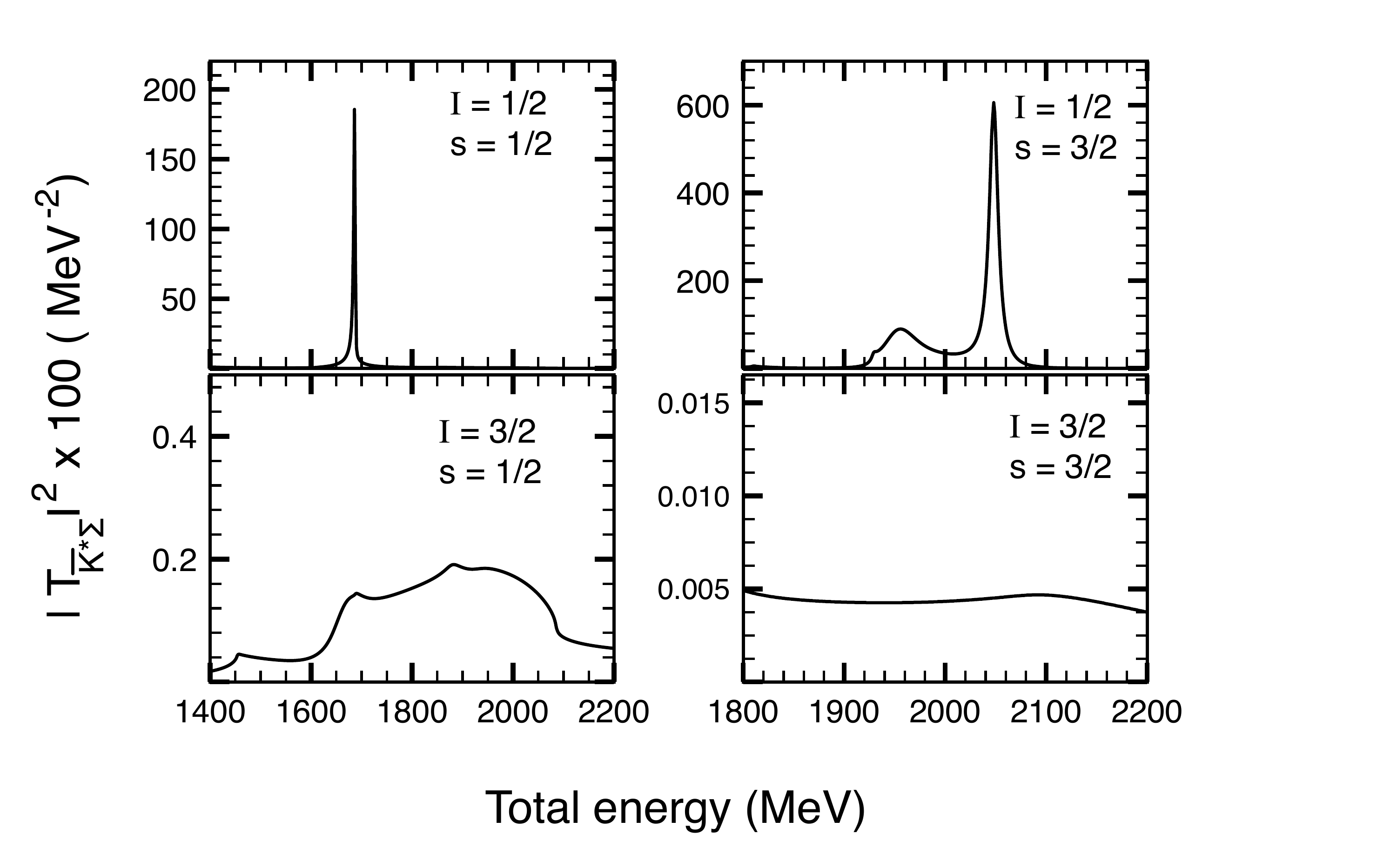}
\caption{Squared amplitudes of the $\bar K^* \Sigma$ channel in different ispospin, spin configurations. }\label{results}
\end{center}
\end{figure}

We show the squared amplitude obtained by solving coupled channel equations for the $\bar K^* \Sigma$ channel, as an example, in different isospin-spin configurations in Fig.~\ref{results}.
It can be seen in Fig.~\ref{results} that a peak structure is found around 1690 MeV in the isospin, spin 1/2 case. As discussed in Ref.~\cite{cascades}, this peak corresponds to a pole in the complex plane at $M - i \Gamma/2 = 1687 - i 2 ~{\rm MeV}$. This result is in good agreement with the values $M~=~\left(1684.7 \pm 1.3^{+2.2}_{-1.6} \right)$~MeV, $\Gamma = \left(8.1^{+3.9+1.0}_{-3.5-0.9}\right)$ MeV obtained by the BABAR Collaboration \cite{babar} and  $M  = \left(1688 \pm 2\right)$ MeV, $\Gamma = \left(11 \pm 4\right)$ MeV obtained by the BELLE Collaboration~\cite{belle}. We have also calculated the decay rate of this state to different open channels: $\pi \Xi$, $\bar K \Lambda$ and $\bar K \Sigma$; the branching ratios obtained are 17$\%$, $28.5\%$ and $54.5\%$, respectively. Using these values we get
\begin{equation}
\frac{B(\Xi^0(1690) \to K^- \Sigma^+)}{B(\Xi^0(1690) \to \bar K^0 \Lambda)} = 0.52, \label{ratio}
\end{equation}
which is in excellent agreement with the data from the Belle collaboration: $0.50 \pm 0.26$ \cite{belle}. Yet another finding reported by the Belle collaboration is that a clear signal of $\Xi(1690)$ is not found in the $\pi \Xi$ mass spectrum. The couplings of the isospin, spin 1/2 state given in Ref.~\cite{cascades} (which we list in Table.~\ref{coupXi1690}, for the reader's convenience) for at least the different channels open for decay, also successfully explain this finding. As can be seen from Table.~\ref{coupXi1690}, the coupling for the $\pi \Xi$ channel is nearly zero. The agreement between all our results and the known information from the experimental data shows that our state can be associated with $\Xi(1690)$. The agreement between the two is remarkable, given the fact that we have not fitted any parameters of the model to reproduce the data. The only parameter of the model is the cut-off value required to regularize the loop functions. We have used one unique value for all channels, $\Lambda = 800$ MeV (see Ref.~\cite{cascades} for more details). We have also tested the stability of our results by varying this cut-off value by $\pm 30$ MeV and find that both the real and imaginary part of the pole value change by $\pm 1 $ MeV.  We can definitely summarize that our findings imply that $\Xi(1690)$ can be interpreted
\begin{table}[h!]
\centering
\caption[]{ Couplings of $\Xi(1690)$ to the different channels open for decay.} \label{coupXi1690}
\begin{tabular}{cc}\hline
Channels &Couplings   \\
\hline
$\pi \Xi$              &$-0.1 - i 0.1$ \\
$\eta \Xi$             &$~0.9 + i 0.2$ \\
$\bar K \Sigma$       & $~1.5 + i 0.2$\\
$\bar K \Lambda$          &$-0.3 + i 0.1$  \\\hline
\end{tabular}
\end{table}
as a dynamically generated state. It is also important to mention that the vector meson-baryon channels play an important role in the generation of $\Xi(1690)$: we do not find this state if we consider only pseudoscalar-baryon coupled channels. This latter finding is in consonance with that of Ref.~\cite{ramosXi}, where a pseudoscalar meson-baryon  coupled channel calculation was done for strangeness $-2$.  Another study on $\Xi$'s does find a pole for $\Xi(1690)$ \cite{sekihara} using only pseudoscalar-baryon channels by fitting the parameters of the model to the experimental data. In Ref.~\cite{sekihara}, the loop functions are calculated using the dimensional regularization method and the subtraction constants for each channel are treated as free parameters. Even then a full width of only 1 MeV is found in Ref.~\cite{sekihara}. Some of the subtraction constant values found in Ref.~\cite{sekihara} deviate from the natural values, which implies that an additional effective potential is added to the system (as explained in Ref.~\cite{hyodo}), and this effective potential may account partly for the missing vector-baryon channels.

Further, it can be seen that a peak in the squared amplitude is also seen in the total isospin 1/2, spin 3/2 case. The mass of this state is $\sim$ 2050 MeV and half width at full maximum is $\sim 8$ MeV.  Although the mass of this state is far from the mass of $\Xi(2120)$, we argue in Ref.~\cite{cascades} that our state can be related to  $\Xi(2120)$. One of the reasons is that the width of $\Xi(2120)$ is known to be $< 20 $ MeV~\cite{pdg}. Another reason is that the signal  of the latter, so far, has been found only in the $\bar K \Lambda$ mass spectrum~\cite{pdg}, and if $\Xi(2120)$ is related to the spin 3/2 state found in our work, then $\bar K \Lambda$ would not be an ideal channel to study this resonance. In such a case, if the signal is not very clear, the determination of the mass may not be reliable. As discussed in Ref.~\cite{cascades}, $\pi \Xi$ and $\bar K \Sigma$ channels are better suited to look for this resonance. We have also checked that changing the cut-off  value used in our work, by $\sim 30$ MeV, changes the mass and width of the spin 3/2 state found in our work in the range $(2046 \pm 6) + i (8.2 \pm 2.2)$ MeV.

The squared amplitudes shown for the isospin 3/2 and spins 1/2, 3/2 show that no state is found in these configurations, only cusps corresponding to opening of different channels can be seen. If a state in isospin 3/2 configuration would have appeared, it would have been an exotic pentaquark state. A claim of the existence of a narrow, exotic $\Xi^{--}$ state, with mass $\sim$ 1862 MeV, in the $\pi^- \Xi^-$ mass spectrum was reported in Ref.~\cite{Alt:2003vb}. However, later investigations did not confirm the existence of this state (see Ref.~\cite{Abelev:2014qqa} for more results on this topic).

\section*{Acknowledgements}
K.P.K, A.M.T, F.S.N and M.N gratefully acknowledge the financial support received from FAPESP (under  the
grant number 2012/50984-4). K.P.K and A.M.T are also thankful for the support from CNPq (under the grant numbers 311524/2016-8 and 310759/2016-1). A.H and H.N thank the support from Grants-in-Aid for Scientific Research (Grants No. JP17K05441(C)) and  (Grants No. JP17K05443(C)), respectively.

\end{document}